\documentstyle[floats,psfig,twocolumn,eqsecnum,prb,aps]{revtex}
\begin{document}
\draft

\twocolumn[\hsize\textwidth\columnwidth\hsize\csname
@twocolumnfalse\endcsname

\title{Bound Magnetic Polaron Interactions in Insulating \\
Doped Diluted Magnetic Semiconductors}
\author{Adam C. Durst$^{1,2}$, R. N. Bhatt$^{1}$, and P. A. Wolff$^{2}$}
\address{$^{1}$Department of Electrical Engineering, Princeton University,
Princeton, New Jersey 08544}
\address{$^{2}$Department of Physics, Massachusetts Institute of Technology,
Cambridge, Massachusetts 02139}
\date{\today}

\maketitle

\begin{abstract}
The magnetic behavior of insulating doped diluted magnetic 
semiconductors (DMS) is characterized by the interaction of large 
collective spins known as bound magnetic polarons.  Experimental
measurements of the susceptibility of these materials have suggested that 
the polaron-polaron interaction is ferromagnetic, in contrast to the 
antiferromagnetic carrier-carrier interactions that are characteristic of
nonmagnetic semiconductors.  To explain this behavior, a model has been 
developed in which polarons interact via both the standard direct
carrier-carrier exchange interaction (due to virtual carrier hopping) and an 
indirect carrier-ion-carrier exchange interaction (due to the interactions of 
polarons with magnetic ions in an interstitial region).  Using a variational
procedure, the optimal values of the model parameters were 
determined as a function of temperature.  At temperatures of interest, the 
parameters describing polaron-polaron interactions were found to be 
nearly temperature-independent.  For reasonable values of these constant
parameters, we find that indirect ferromagnetic interactions
can dominate the direct antiferromagnetic interactions and cause the
polarons to align.  This result supports the experimental evidence for
ferromagnetism in insulating doped DMS.
\end{abstract}

]

\section{Introduction}
\label{sec:intro}
Diluted magnetic semiconductors (DMS) are semiconductors in which a
fraction of the nonmagnetic ions that make up the crystal structure
have been replaced by magnetic transition metal or rare earth ions.
For example, substituting Mn$^{2+}$ ions for some of the Cd ions
in the nonmagnetic semiconductor, CdTe, yields the diluted magnetic
semiconductor, Cd$_{1-x}$Mn$_{x}$Te.  In doped DMS, the sizable exchange interaction
between magnetic ions and carriers (electrons or holes) leads to
unusual optical, magnetic, and transport properties.  Due to their
potential for use in novel devices which take advantage of both
their semiconducting and magnetic properties, DMS have, of late, been
the subject of much interest. \cite{ohn98,pri98}
Recently, the discovery \cite{mat98} of a ferromagnetic transition temperature
of 110~K in a sample of Ga$_{1-x}$Mn$_{x}$As with $x \approx 0.05$
has further enhanced both the
experimental \cite{die01,bes99,ohn99,szc99,die00} and
theoretical \cite{kon00,sch01,bha99,wan00,bha00,ang00,ken01} interest in DMS.

\begin{figure}
\centerline{\psfig{file=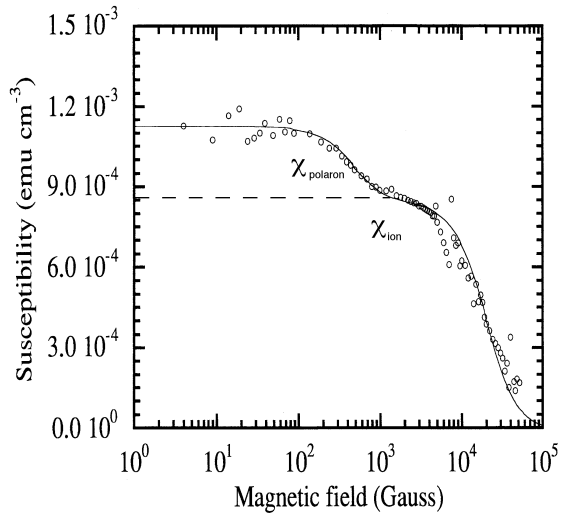}}
\vspace{0.5cm}
\caption{Magnetic susceptibility ($dM/dH$) of Zn$_{1-x}$Mn$_{x}$Te ($x \approx 0.1$),
p-doped at the level of $3 \times 10^{17} / \mbox{cm}^{3}$, measured
at 2~K as a function of applied magnetic field.  Circles denote measured data
while the solid line is a fit to a dual magnetization model.  The dashed line
separates the collective spin (polaron) contribution from the individual Mn
spin contribution.  Data was obtained by J. Liu \cite{liu94} at NEC Research
Institute and is reproduced here with the permission of NEC.}
\label{fig:doublestep}
\end{figure}

In the II-VI DMS Zn$_{1-x}$Mn$_{x}$Te ($x \le 0.1$), p-doped with carriers
at the level of $3 \times 10^{17} / \mbox{cm}^{3}$, where the system is in the
insulating state, measurements of susceptibility versus applied magnetic field
were conducted by J. Liu at NEC Research Institute.  The data, originally
reported in an NEC technical memo \cite{liu94}, has been reproduced in
Fig.~\ref{fig:doublestep} for the convenience of the reader.  Note the
double-step structure of the susceptibility and the two characteristic field
scales indicated by the inflection points of the curve.  This form suggests
a dual magnetization mechanism whereby large collective spins
align at fields ($\sim 300$~G) too weak to magnetize the individual magnetic ions.
Only at much larger fields ($\sim 15,000$~G) do the individual Mn spins align.
Within this interpretation, the measured susceptibility can be viewed as the
sum of two contributions: a collective spin term that drops off around 300~G
and an individual spin term that drops off around 15,000~G.  The dashed line
in Fig.~\ref{fig:doublestep} serves to separate these two contributions.

For the $x$-regime in question, the Mn concentration is not enough to percolate,
and the undoped system is not magnetically ordered.  (Spin glass type order of the
undoped system has been observed in II-VI DMS for Mn concentrations above
$x = 0.2$. \cite{wol88})  Consequently, the unusual magnetic behavior is
attributable to the presence of the dopants.  This is not surprising,
despite the low carrier concentration, because the Bohr radius which
characterizes the carrier wave function is large compared to the Mn 3d
wave function which characterizes the extent of the Mn local moment.
We interpret the large collective spins, responsible for the double-step
form of the susceptibility, to be bound magnetic polarons,
formed by the exchange interaction between localized carriers
and magnetic ions within the carrier orbit.
Furthermore, a fit of the polaron part of the
the susceptibility data to a Curie-Weiss form reveals a net ferromagnetic
interaction between the polarons.  This result is in stark contrast to
that observed for conventional nonmagnetic semiconductors in which
virtual carrier hopping invariably yields antiferromagnetism. \cite{bha82}
To explain both the formation of bound magnetic polarons and the
ferromagnetic nature of their interaction, we introduced,
in Ref.~\onlinecite{wol96}, a bound magnetic polaron model for insulating
doped DMS.  This was further elucidated by a comprehensive calculation
in which we showed how the parameters of the model could be obtained in
an optimal manner using a variational principle \cite{dur95}, which
we present below.

In Sec.~\ref{sec:polaronpair}, we describe the system of two interacting
polarons in a diluted magnetic semiconductor and develop a simplified
model to capture its behavior.  In Sec.~\ref{sec:partitionfunction},
we calculate, within our model, both the single polaron partition function
and the polaron-pair partition function.  Making use of these partition
functions, we implement a variational procedure, in Sec.~\ref{sec:variational},
to optimize the parameters of our model.  We find that while the model parameters
describing single polaron formation are temperature-dependent, the
polaron-polaron interaction parameters can be treated as temperature-independent
constants for magnetic ion and carrier densities of interest.  We make use of
these results in Sec.~\ref{sec:cosine} where we demonstrate how a
ferromagnetic polaron-polaron interaction can come about.  Conclusions
are presented in Sec.~\ref{sec:conclusions}.

\section{Polaron-Pair System and Model}
\label{sec:polaronpair}
\subsection{The System}
\label{ssec:system}
To understand the magnetic behavior of diluted magnetic 
semiconductors, we consider the polaron-pair system which consists of two 
carriers (electrons or holes) bound to impurity sites (donors or acceptors) 
separated by an inter-impurity distance, $R_{12}$, and the magnetic ions (usually spin 
5/2 Mn) which surround them.  This complex system interacts via three 
independent exchange interactions each of which have different characteristic 
length scales.

The bound carriers interact directly via an impurity-impurity exchange 
interaction.  Although this interaction can be more complicated for the case of 
acceptors (for which the valence band is degenerate) \cite{dur96},
we assume an
impurity-impurity interaction of the Heisenberg type, as in the donor case,
characterized by an exchange constant $J$.  This interaction has been shown to be 
antiferromagnetic for donors in nonmagnetic semiconductors \cite{bha82} and is
assumed to be so for carriers in diluted magnetic semiconductors as well.  $J$ is a
function of the inter-impurity distance and the effective Bohr radius and varies as
\begin{equation}
J \sim \exp (-2 R_{12}/a_{B}) .
\label{eq:Jdef}
\end{equation}
Were it not for the influence of additional interactions, this direct exchange 
would yield a net antiferromagnetic exchange interaction in DMS.

The second interaction at work in this system is the exchange interaction 
between each of the carriers and the magnetic ions.  This interaction is 
also antiferromagnetic and is proportional in magnitude to $\alpha|\Psi|^{2}$
where $\alpha$ is the carrier-ion exchange constant for the particular material
and $\Psi$ is the carrier wave function.  For the purpose of this study, we will
take $\Psi$ to be the hydrogenic wave function
\begin{equation}
\Psi({\bf r}) = (\pi a_{B}^{3})^{-1/2} \exp (-r/a_{B})
\label{eq:hydrowavefunc}
\end{equation}
with an effective Bohr radius $a_{B}$.  However, it should be noted that for
acceptors in particular, the carrier wave functions may be more complicated.
\cite{bal73}  We are also implicitly assuming that the binding energy of the
impurity is large compared to the magnetic energy of the polaron, so magnetic
ordering does not change the carrier wave function.

Finally, there exists an additional antiferromagnetic exchange interaction,
between the individual magnetic ions, which has a characteristic length scale on 
the order of a magnetic ion radius ($\sim \mbox{\AA}$).  Since this length scale is
small compared with others in the system (i.e.\ $a_{B} \sim 10-20 \mbox{\AA}$), we
neglect all but nearest neighbor interactions and assume that the nearest
neighbors form inert singlets.  Thus, ion-ion interactions are considered only via
the use of an effective magnetic ion concentration, $\bar{x} \equiv x(1-x)^{12}$,
in place of $x$, the true magnetic ion concentration. \cite{wol88}

\begin{figure}
\centerline{\psfig{file=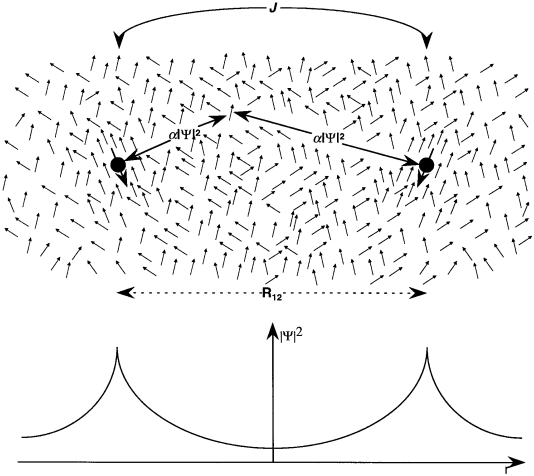}}
\vspace{0.5cm}
\caption{Schematic of polaron-pair system.}
\label{fig:system}
\end{figure}

Hence, the polaron-pair system (given the assumptions noted above) interacts via
two antiferromagnetic exchange interactions, a carrier-carrier interaction and
a carrier-magnetic ion interaction.  These interactions are depicted in
Fig.~\ref{fig:system} and result in the Hamiltonian
\begin{equation}
H = \alpha \sum_{n} {\bf s}_{1} \cdot {\bf S}_{n} | \Psi_{1n} |^{2}
+ \alpha \sum_{n} {\bf s}_{2} \cdot {\bf S}_{n} | \Psi_{2n} |^{2}
+ J {\bf s}_{1} \cdot {\bf s}_{2}
\label{eq:trueHamiltonian}
\end{equation}
where $n$ runs over all magnetic ions, $\Psi_{1n}$ and $\Psi_{2n}$ are the carrier
wave functions at the magnetic site $n$, $s_{1}$ and $s_{2}$ are the carrier
spins, and the $S_{n}$ are magnetic ion spins.

We consider this to be a polaron-pair system because carrier-ion interactions 
tend to anti-align the spins of magnetic ions in the vicinity of a carrier with 
respect to the carrier spin.  Thus, each carrier and the ions in its vicinity form a 
single magnetic polaron with a large collective spin.  The polarons interact via 
both the direct antiferromagnetic carrier-carrier exchange interaction and the 
indirect ferromagnetic exchange interaction that results when carrier-ion 
interactions cause both polarons to anti-align with the same magnetic ions.
Details of the competition between these two interactions will be explored as
we study the nature of a simplified model.

\subsection{The Model}
\label{model}
Although the true Hamiltonian provides the best description of the 
polaron-pair system, its solution is complicated by the fact that the magnitude of 
the carrier-ion interaction varies exponentially with carrier-ion distance (since 
carrier wave functions are hydrogenic).  In order to obtain a more detailed 
understanding of the polaron-pair system, it is necessary to study a simplified, 
more tractable model.  Such a model is obtained by making two simplifying 
approximations: the single step approximation and the interstitial region 
approximation.

The single step approximation entails replacing the carrier wave functions 
by radial step functions that are constant up to a radius $R$ and zero beyond $R$.
In this approximation, all of the magnetic ions within a sphere of radius $R$ about
a carrier interact with that carrier with the same exchange constant, $K$.  Thus, in 
this model, the definition of a polaron becomes clear.  A polaron is composed of 
a single carrier and all of the magnetic ions within a radius $R$ of the impurity
site to which the carrier is bound.  This approximation, first developed by 
Ryabchenko and Semenov \cite{rya83}, allows for the exact calculation of the single 
polaron partition function and makes the polaron-pair case much more tractable.

To consider interactions between two polarons, we must make 
the additional conjecture that there is an interstitial region between the two 
polarons within which the magnetic ions interact significantly with both carriers.  
Such a region must exist in order for the indirect ferromagnetic
carrier-ion-carrier interactions to be significant.  In order to treat the
effects of these interstitial ions within our model, we assume a cylindrically
symmetric interstitial region within which all of the magnetic ions interact with
both of the carriers with an exchange constant, $K^{\prime}$.  In this
interstitial region approximation \cite{bha95}, carrier-ion exchange causes
both carrier spins to anti-align with the interstitial spins and thereby align
with each other.  Thus, an indirect source of carrier-carrier ferromagnetism is
introduced into the model.

\begin{figure}
\centerline{\psfig{file=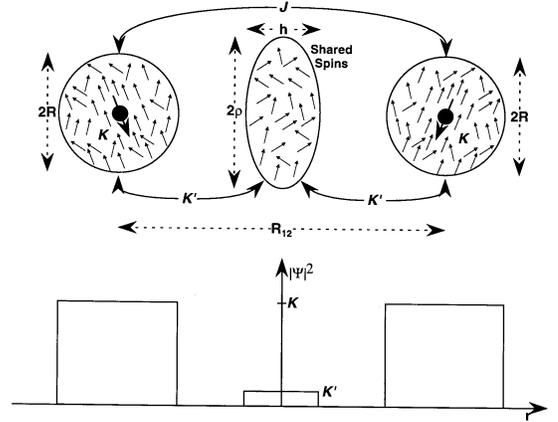}}
\vspace{0.5cm}
\caption{Schematic of polaron-pair model.}
\label{fig:model}
\end{figure}

In the end, the above approximations yield the model Hamiltonian
\begin{mathletters}
\label{eq:modelHamiltonian}
\begin{eqnarray}
H_{m} &=& K \left[ ({\bf s}_{1} \cdot {\bf S}_{1})
+ ({\bf s}_{2} \cdot {\bf S}_{2}) \right] \nonumber \\
&+& K^{\prime} ({\bf s}_{1} + {\bf s}_{2}) \cdot {\bf S}_{3}
+ J {\bf s}_{1} \cdot {\bf s}_{2}
\end{eqnarray}
\begin{equation}
{\bf S}_{1} \equiv \sum_{\mbox{\tiny Sphere \# 1}} {\bf S}_{i} \;\;\;\;\;\;
{\bf S}_{2} \equiv \sum_{\mbox{\tiny Sphere \# 2}} {\bf S}_{j} \;\;\;\;\;\;
{\bf S}_{3} \equiv \sum_{\mbox{\tiny Interstitial}} {\bf S}_{k}
\end{equation}
\end{mathletters}
where $K$ is the intra-polaron ion-carrier exchange constant, $K^{\prime}$ is
the interstitial ion-carrier exchange constant, $J$ is the direct carrier-carrier
exchange constant, ${\bf s}_{1}$ and ${\bf s}_{2}$ are the carrier spins,
${\bf S}_{1}$ and ${\bf S}_{2}$ are the net polaron spins, and ${\bf S}_{3}$ is the
collective spin of the interstitial region.  (At this point, we specify only that
the interstitial region have cylindrical symmetry and be located between the 
polarons.  However, for computational purposes, a particular shape must be 
chosen.  This matter is discussed further in Sec.~\ref{ssec:varpolpair}.)
As is indicated in Fig.~\ref{fig:model} where the details of this model are
presented graphically, the essence of the 
polaron-polaron model reduces to a competition between the direct 
antiferromagnetic carrier-carrier interactions characterized by $J$, and the indirect 
ferromagnetic carrier-ion-carrier interactions characterized by $K^{\prime}$.
By showing that there are circumstances in which the ferromagnetic interaction
dominates, a theoretical justification for DMS ferromagnetism can be obtained.
By applying the RS (Ryabchenko and Semenov) single step approximation 
and the interstitial approximation to the polaron-pair model, we have effectively 
separated the polaron-pair system into one mechanism for polaron formation 
and another for polaron-polaron interaction.  It is this separation that makes 
possible a calculation of the polaron-pair partition function.

\section{Partition Function Calculation}
\label{sec:partitionfunction}
To obtain the partition function for this problem, we must consider both
the interacting part of the system (the carriers and magnetic ions 
within the polarons and the interstitial region), and the non-interacting part 
(those magnetic ions that are external to both the polarons and the interstitial 
region).  Thus, the full partition function takes the form,
$Z = Z_{pp} Z_{ext}$, where $Z_{pp}$ is the polaron-pair partition function
and $Z_{ext}$ is the partition function of the non-interacting external ion spins.

Since the external ions are inert, $Z_{ext}$ is just given by the degeneracy of
the magnetic ion spins.  Taking the magnetic ions to be spin-5/2 (as is the case
for Mn), each spin has $2s+1=6$ orientations.  Therefore, $Z_{ext}=6^{N_{ext}}$,
where $N_{ext}$, the number of external magnetic ions in the system, is equal to the 
total number of ions in the system minus the number of ions in the polarons and 
interstitial region.  Note that while these non-interacting external spins
contribute no energy to the system, they contribute nonzero entropy and therefore
cannot be neglected.

Due to the approximations made in developing our model, the polaron-pair
system separates into an individual polaron part and a polaron-polaron
interaction part.  Hence, as will be explicitly shown, the polaron-pair
partition function can be expressed as the product of two single polaron
partition functions and a polaron-polaron interaction partition function.

\subsection{Single Polaron Partition Function}
\label{ssec:partfuncsingle}
The single polaron partition function can be calculated exactly for the 
single step model that we have adopted.  For a single polaron, the Hamiltonian is
\begin{equation}
H = \lambda_{s} K {\bf s} \cdot {\bf S} = \frac{\lambda_{s} K}{2}
\left[ ({\bf s} + {\bf S})^{2} - {\bf s}^{2} - {\bf S}^{2} \right]
\label{eq:Hsinglepol}
\end{equation}
where ${\bf s}$ is the carrier spin, ${\bf S}$ is the sum of the magnetic
ion spins within the polaron, and $\lambda_{s}$ is a placeholder constant
that has been inserted for notational convenience and will eventually be set
equal to one.  For a given $S$, the total spin can take
two values: $S+\frac{1}{2}$ or $S-\frac{1}{2}$.  The former yields
an energy and degeneracy
\begin{equation}
E_{+} = \frac{\lambda_{s} KS}{2} \;\;\;\;\;\; g_{+} = 2(S+1)
\label{eq:Eplus}
\end{equation}
while the latter yields
\begin{equation}
E_{-} = -\frac{\lambda_{s} K(S+1)}{2} \;\;\;\;\;\; g_{-} = 2S .
\label{eq:Eminus}
\end{equation}
Thus, the single polaron partition function is given by
\begin{equation}
Z_{pol} = \mbox{Tr} \left[ e^{-\beta H} \right] = \sum_{S} D(S)
\left[ g_{+} e^{-\beta E_{+}} + g_{-} e^{-\beta E_{-}} \right]
\label{eq:singlepol1}
\end{equation}
where $D(S)$ is the number of ways in which the ion spins can be arranged to give 
a collective spin $S$.
Defining $D_{z}(S)$ to be the number of ways that the ions can be arranged to give 
a collective z-component of spin equal to $S$ and doing a bit of algebraic 
manipulation, we obtain
\begin{equation}
Z_{pol} = 2 \left[ 1+ \left( e^{\gamma}-1 \right) \frac{\partial}{\partial\gamma}
\right] \sum_{S=-5N_{1}/2}^{5N_{1}/2} D_{z}(S) \cosh (\gamma S)
\label{eq:singlepol2}
\end{equation}
where $\gamma \equiv \lambda_{s}\beta K/2$ and $N_{1}$ is the
number of ion spins within the polaron.
Using the definition of a delta-function, $D_{z}(S)$ can be written as
\begin{equation}
D_{z}(S) = \mbox{Tr}\ \delta \left( S - \sum_{j=1}^{N_{1}} S^{z}_{j} \right)
= \int_{-\infty}^{\infty} \frac{d\lambda}{2\pi} e^{i\lambda S}
\left[ 6 \, F(i\lambda) \right]^{N_{1}}
\label{eq:Dzdef}
\end{equation}
where
\begin{equation}
F(x) = \frac{1}{6} \left[ e^{\frac{5x}{2}} + e^{\frac{3x}{2}} + e^{\frac{x}{2}}
+ e^{-\frac{x}{2}} + e^{-\frac{3x}{2}} + e^{-\frac{5x}{2}} \right] .
\label{eq:Fdef}
\end{equation}
For large $N_{1}$, the sum over $S$ can be converted to an integral and the partition 
function can be written as
\begin{eqnarray}
Z_{pol} &=& 2 \left[ 1+ \left( e^{\gamma}-1 \right) \frac{\partial}{\partial\gamma}
\right] \int_{-\infty}^{\infty} dS \nonumber \\
&\times& \int_{-\infty}^{\infty} \frac{d\lambda}{2\pi}
\left( e^{\gamma S} + e^{-\gamma S} \right) e^{i\lambda S}
\left[ 6 \, F(i\lambda) \right]^{N_{1}} .
\label{eq:singlepol3}
\end{eqnarray}
Continuing to the imaginary temperature axis, using the definition of
a delta-function, and continuing back to the real temperature axis, we obtain 
\begin{equation}
Z_{pol} = 6^{N_{1}} Z_{1} \;\;\;\;\;\;
Z_{1} = 2 \left[ 1+ \left( e^{\gamma}-1 \right)
\frac{\partial}{\partial\gamma}\right] F(\gamma)^{N_{1}}
\label{eq:singlepol4}
\end{equation}
where we have separated out the factor of $6^{N_{1}}$ which will be
cancelled by part of $Z_{ext}$ in the full partition function.
Note that this expression has the correct infinite temperature limit since
$Z_{pol}(T \rightarrow \infty) \rightarrow 2(6)^{N_{1}}$, which is the
partition function for a non-interacting system of $N_{1}$ spin-5/2
magnetic ions and one spin-1/2 carrier.

\subsection{Polaron-Pair Partition Function}
\label{ssec:partfuncpair}
The exact quantum mechanical 
calculation of the polaron-pair partition function is significantly more
complicated than the single polaron case.  However, at 
low temperatures, a semiclassical technique introduced in
Ref.~\onlinecite{wol96} can be used to 
find $Z_{pp}$ throughout the temperature range of interest.  Specifically, we must 
make the assumption that the temperature is low enough such that the ion spins 
within the polarons and interstitial region are well enough aligned that the two 
polaron spins, $S_{1}$ and $S_{2}$, and the interstitial region spin, $S_{3}$,
are large enough to 
be treated as classical magnetic moments.  Thus, we make a semiclassical 
approximation in which $S_{1}$, $S_{2}$, and $S_{3}$ are treated as
classical spins while the 
carrier spins, $s_{1}$ and $s_{2}$, are treated quantum mechanically.
For the case we are interested in, appropriate for light to moderately doped
II-VI based DMS, we expect that $K^{\prime},J \ll K$.
Consequently, we will first find the partition
function for the case of non-interacting polarons ($K^{\prime}=J=0$) and then
include the effects of nonzero $K^{\prime}$ and $J$ as 
first order perturbations.  Separating the Hamiltonian into three parts, we write
$H = H_{0} + H_{1} + H_{2}$ where
\begin{eqnarray}
H_{0} &=& K \left[ \lambda_{s1} ({\bf s}_{1} \cdot {\bf S}_{1})
+ \lambda_{s2} ({\bf s}_{2} \cdot {\bf S}_{2}) \right] \nonumber \\
H_{1} &=& J ({\bf s}_{1} \cdot {\bf s}_{2}) \nonumber \\
H_{2} &=& K^{\prime} \left[ \lambda_{c1} ({\bf s}_{1} \cdot {\bf S}_{3})
+ \lambda_{c2} ({\bf s}_{2} \cdot {\bf S}_{3}) \right]
\label{eq:H0H1H2}
\end{eqnarray}
and we have introduced four new constants, $\lambda_{s1}$, $\lambda_{s2}$,
$\lambda_{c1}$, and $\lambda_{c1}$.  While these constants will be set equal
to unity at the end of our calculation, they act as placeholders which will
be useful when we optimize our model parameters in Sec.~\ref{sec:variational}.

In the non-interacting polarons limit ($K^{\prime}=J=0$), the polaron-pair
partition function is simply $Z_{pp}=6^{N_{1}+N_{2}}Z_{1}Z_{2}$
where $Z_{1}$ and $Z_{2}$
are the single polaron partition functions given by 
Eq.~\ref{eq:singlepol4} with $\gamma_{1}=\lambda_{s1}\beta K/2$ and
$\gamma_{2}=\lambda_{s2}\beta K/2$ respectively.  In the semiclassical limit,
the wave functions of the non-interacting polaron-pair are:
$\alpha(1)\alpha(2)$, $\alpha(1)\beta(2)$, $\beta(1)\alpha(2)$, and
$\beta(1)\beta(2)$, where $\alpha$ is the total spin $S+1/2$ state of the
single polaron (carrier and magnetic ions aligned)
and $\beta$ is the $S-1/2$ state (carrier and magnetic ions anti-aligned).
At the low temperatures for which the 
semiclassical approximation is valid, only the ground state,
$\beta(1)\beta(2)$, is significantly occupied.  Hence, the $K^{\prime}$
and $J$ perturbations are taken to be the diagonal matrix elements in this
ground state.

Therefore, for interacting polarons with nonzero $K^{\prime}$ and $J$, we write
\begin{equation}
M_{J} = \langle \beta(1)\beta(2) | H_{1} | \beta(1)\beta(2) \rangle
= \frac{J}{4} \frac{{\bf S}_{1} \cdot {\bf S}_{2}}{S_{1} S_{2}}
= \frac{J \mu_{12}}{4}
\label{eq:MJ}
\end{equation}
\begin{eqnarray}
M_{K^{\prime}} &=& \langle \beta(1)\beta(2) | H_{2} | \beta(1)\beta(2) \rangle
\nonumber \\
&=& -\frac{K^{\prime}}{2} \left[ \frac{\lambda_{c1} S_{2} {\bf S}_{1}
+ \lambda_{c2} S_{1} {\bf S}_{2}}{S_{1} S_{2}} \cdot {\bf S}_{3} \right]
= -\frac{\Omega}{\beta} \mu_{3} {\bf S}_{3} \nonumber \\ &&
\label{eq:MKprime}
\end{eqnarray}
where
\begin{equation}
\Omega = \frac{\beta K^{\prime}}{\sqrt{2}}
\sqrt{\frac{\lambda_{c1}^{2} + \lambda_{c2}^{2}}{2}
+ \lambda_{c1} \lambda_{c2} \mu_{12}} ,
\label{eq:Omegadef}
\end{equation}
$\mu_{12}$ is the cosine of the angle between $S_{1}$ and $S_{2}$,
and $\mu_{3}$ is the cosine of the angle 
between $S_{3}$ and the z-axis.  Making use of these matrix elements,
the polaron-pair partition function can be written as
\begin{eqnarray}
Z_{pp} &=& \mbox{Tr} \left[ e^{-\beta (H_{0} + M_{J} + M_{K^{\prime}})} \right]
= \int d^{3}S_{1} d^{3}S_{2} d^{3}S_{3} \nonumber \\
&\times& D(S_{1}) D(S_{2}) D(S_{3})
e^{\gamma_{1}S_{1} + \gamma_{2}S_{2} - \frac{\beta J}{4} + \Omega \mu_{3} S_{3}} .
\nonumber \\ &&
\label{eq:polpair1}
\end{eqnarray}
Performing the indicated integration and doing a bit of algebra this becomes
\begin{equation}
Z_{pp} = 6^{N_{1}+N_{2}} Z_{1} Z_{2}
\int_{0}^{\sqrt{2}} e^{-\frac{\beta J (x^{2}-1)}{4}}
\left[ 6\, F(\Omega) \right]^{N_{3}} x\, dx
\label{eq:polpair2}
\end{equation}
where $x^{2}=1+\mu_{12}$, $N_{3}$ is the number of magnetic ion spins within
the interstitial region, and we have identified $Z_{1}$ and $Z_{2}$ as the
single polaron partition 
functions.  Notice that in this approximation, the partition function does separate 
into a polaron formation factor and a polaron-polaron interaction factor.
Multiplying this result by the partition function for external ion spins,
$Z_{ext}=6^{N_{tot}-N_{1}-N_{2}-N_{3}}$, and dropping the constant factor
of $6^{N_{tot}}$, we obtain the full partition function
\begin{equation}
Z = Z_{1} Z_{2} Z_{3} \;\;\;\;\;\;
Z_{3} = \int_{0}^{\sqrt{2}} e^{-\frac{\beta J (x^{2}-1)}{4}}
F(\Omega)^{N_{3}} x\, dx
\label{eq:polpair3}
\end{equation}
where $Z_{3}$ is the polaron-polaron interaction part.

\section{Variational Optimization of Model Parameters}
\label{sec:variational}
Given the partition function calculated in the previous section, we proceed
to optimize the parameters of our model via a variational approach.  At
$T=0$, optimal values of the model parameters could be obtained by
minimizing the expectation value of the model Hamiltonian.  At $T \neq 0$,
we adopt an analogous variational approach described by Feynman \cite{fey72}
for which the quantity to be minimized is the ${\cal F}$-function
\begin{equation}
{\cal F} \equiv F_{m} + \langle H - H_{m} \rangle
\label{Ffunc}
\end{equation}
where $H_{m}$ is the model Hamiltonian and $H$ is the true hamiltonian.
The average $\langle \ldots \rangle$ is taken over the states of $H_{m}$.
${\cal F}$ can be shown \cite{fey72}
to be an upper bound on the true free energy $F$
of the Hamiltonian $H$ at the temperature T in question.  By minimizing
${\cal F}$ with respect to the model parameters, optimal values can be
determined as a function of temperature.

As will be shown explicitly in Sec.~\ref{ssec:varpolpair}, the total
${\cal F}$-function separates into the sum of two single polaron
functions and a polaron-polaron interaction function.  Therefore,
we shall optimize the single polaron parameters first and then consider
the interaction parameters.

\subsection{Single Polaron Parameter Optimization}
\label{ssec:varsinglepolaron}
The single polaron ${\cal F}$-function can be obtained by expressing
$\langle H_{m} \rangle$, $\langle H \rangle$, and $F_{m}$
in terms of the single polaron partition function, $Z_{1}$.  Recall
that the model Hamiltonian has the form
\begin{equation}
H_{m} = \lambda_{s} K \sum_{j} {\bf s} \cdot {\bf S}_{j}
\label{eq:singleHm}
\end{equation}
where index $j$ runs over all magnetic ion spins, ${\bf S}_{j}$, within a sphere
of radius $R$ about the carrier spin, ${\bf s}$, and $\lambda_{s}$ is a constant
which will soon be set equal to unity.  Taking the thermal average
over the eigenstates of $H_{m}$ yields that
\begin{equation}
\langle H_{m} \rangle = \lambda_{s} K N_{1}
\langle {\bf s} \cdot {\bf S}_{j} \rangle = -\frac{\lambda_{s}}{\beta}
\frac{\partial \ln Z_{1}}{\partial \lambda_{s}}
\label{eq:singleaveHm}
\end{equation}
where $N_{1}$ is the number of magnetic ions within the polaron.
The true Hamiltonian has the form
\begin{equation}
H = \alpha \sum_{n} {\bf s} \cdot {\bf S}_{n} | \Psi_{n} |^{2}
\label{eq:singleH}
\end{equation}
where the index $n$ runs over all magnetic ion spins.  Therefore, noting that
$\langle {\bf s} \cdot {\bf S}_{n} \rangle$ is only nonzero for ion spins
within the polaron, we find that
\begin{equation}
\langle H \rangle = \alpha \sum_{j} | \Psi_{j} |^{2}
\langle {\bf s} \cdot {\bf S}_{j} \rangle = -\frac{\gamma_{s}}{\beta}
\frac{\partial \ln Z_{1}}{\partial \lambda_{s}}
\label{eq:singleaveH}
\end{equation}
where
\begin{equation}
\gamma_{s} = \frac{\alpha}{K V_{s}} \int_{S} d^{3}r\, | \Psi(r) |^2
\label{eq:gammasdef}
\end{equation}
and the integral is over a sphere of radius $R$ and volume $V_{s}$.
Finally, since the free energy is
\begin{equation}
F_{m} = -\frac{1}{\beta} \ln Z_{1}
\label{eq:singleFm}
\end{equation}
we can combine the expressions above (setting $\lambda_{s}=1$) to obtain
the single polaron ${\cal F}$-function
\begin{equation}
{\cal F}_{1} = -\frac{1}{\beta} \left[ \ln Z_{1} + (\gamma_{s} - 1)
\frac{\partial \ln Z_{1}}{\partial \lambda_{s}} \right]
\label{eq:singleFfunc}
\end{equation}
where $Z_{1}$ is given by Eq.~(\ref{eq:singlepol4}).
By minimizing ${\cal F}_{1}$ with respect to the parameters $R$ and $K$,
the optimal values of these parameters can be found.

\begin{figure}
\centerline{\psfig{file=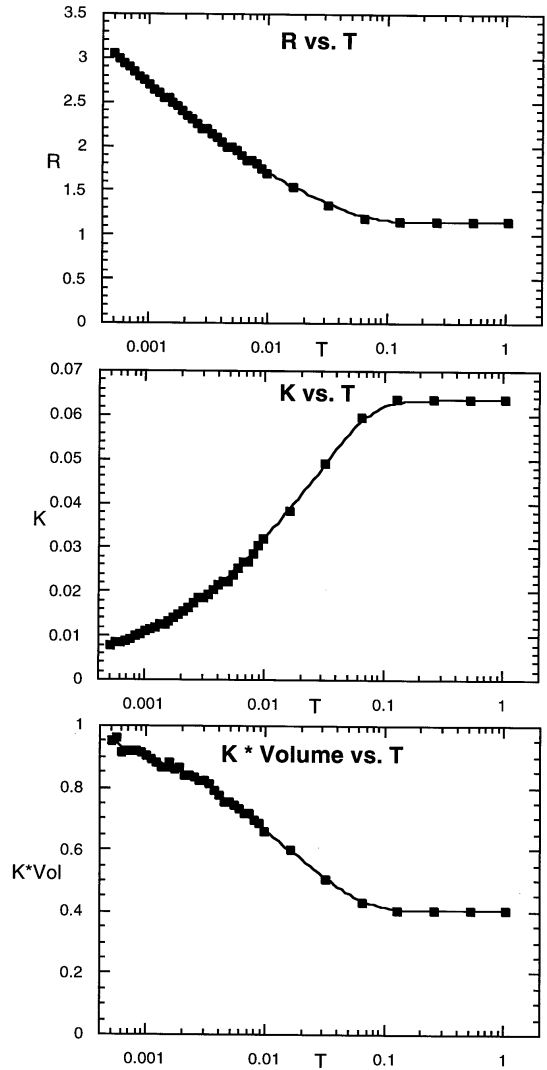}}
\vspace{0.5cm}
\caption{Temperature dependence of single polaron parameters.  All distances
are in units of $a_{B}$ and all energies are in units of
$\alpha/a_{B}^{3} \sim 625$~K.}
\label{fig:singlepolparams}
\end{figure}

Performing such a procedure numerically over a range of temperature
values, the optimal values of the single polaron parameters
were determined as functions of temperature.  The results of this
optimization for a magnetic ion density of 5 ions per sphere of
radius $1a_{B}$ are plotted in Fig.~\ref{fig:singlepolparams}.

In the high temperature limit ($T \gg K$), the exchange interaction between 
the carrier and the magnetic ions within the polaron is insignificant compared to 
temperature.  Thus, the magnetic ion spins are not aligned and there is no 
difference between the free energy of a spin within the polaron and that of an 
external spin.  As a result, the ${\cal F}$-function is minimized when the model 
carrier wave function best matches the true carrier wave function.  This matching 
of a step of width $R$ and height $K$ to a hydrogenic wave function yields the 
optimal, temperature-independent values of $R$ and $K$.  Thus, as is shown in
Fig.~\ref{fig:singlepolparams}, $R$, $K$, and $KV_{s}$ (the total exchange energy)
are temperature-independent in the high temperature regime.

For low temperatures ($T \ll K$), the carrier-ion exchange interaction is 
significant compared to temperature.  Thus, the magnetic ion spins located near 
the carrier anti-align with the carrier spin.  In this situation, the inclusion of an 
additional ion within the polaron entails a gain in exchange energy.  However, 
since the number of external spins decreases by one, there is also a decrease in 
the entropy of free spins.  Therefore, the optimal $R$ is determined by the balance 
of exchange energy and the entropy of free spins.  As $T$ decreases, the exchange 
energy gained by increasing the size of the polaron becomes more valuable.  
Thus, as is shown in Fig.~\ref{fig:singlepolparams}, $R$ increases as
$\log (1/T)$ as $T$ drops.  As $R$
increases, $K$ must decrease in order to maintain the match between the model 
step wave function and the true wave function.  Thus, $K$ decreases with 
decreasing $T$.  Finally, despite the decrease in $K$, the total exchange energy 
(which is proportional to $KV_{s}$) increases as $T$ drops and the spins align.

It is interesting to note that in this problem, the variational principle
leads one to match the Hamiltonian at {\it high} $T$, while entropy-energy
balance determines the parameters at {\it low} $T$.  This is just the
converse of what one expects in most problems.

\subsection{Polaron-Pair Parameter Optimization}
\label{ssec:varpolpair}
Just as for the single polaron, we can obtain the polaron-pair
${\cal F}$-function by expressing
$\langle H_{m} \rangle$, $\langle H \rangle$, and $F_{m}$
in terms of the full partition function, $Z$.
For the polaron-pair, our model Hamiltonian is
\begin{eqnarray}
H_{m} &=& \lambda_{s1} K \sum_{i} {\bf s}_{1} \cdot {\bf S}_{i}
+ \lambda_{s2} K \sum_{j} {\bf s}_{2} \cdot {\bf S}_{j} \nonumber \\
&+& \lambda_{c1} K^{\prime} \sum_{k} {\bf s}_{1} \cdot {\bf S}_{k}
+ \lambda_{c2} K^{\prime} \sum_{k} {\bf s}_{2} \cdot {\bf S}_{k}
+ J {\bf s}_{1} \cdot {\bf s}_{2} \nonumber \\ &&
\label{eq:pairHm}
\end{eqnarray}
where indices $i$ and $j$ run over the magnetic ions in polaron 1 and 2
respectively, $k$ runs over ions in the cylindrically symmetric interstitial
region, and the $\lambda$'s are constants which will soon be set equal to one.
Taking the thermal average then yields
\begin{eqnarray}
\langle H_{m} \rangle
&=& \lambda_{s1} K N_{1} \langle {\bf s}_{1} \cdot {\bf S}_{i} \rangle
+ \lambda_{s2} K N_{2} \langle {\bf s}_{2} \cdot {\bf S}_{j} \rangle
\nonumber \\
&+& \lambda_{c1} K^{\prime} N_{3} \langle {\bf s}_{1} \cdot {\bf S}_{k} \rangle
+ \lambda_{c2} K^{\prime} N_{3} \langle {\bf s}_{2} \cdot {\bf S}_{k} \rangle
\nonumber \\
&+& J \langle {\bf s}_{1} \cdot {\bf s}_{2} \rangle
\label{eq:pairaveHm1}
\end{eqnarray}
where $N_{1}=N_{2}$ are the number of ions in polaron
1 and 2, and $N_{3}$ is the number of ions in the interstitial region.
Noting that
\begin{eqnarray}
\langle {\bf s}_{1} \cdot {\bf S}_{i} \rangle
&=& \langle {\bf s}_{2} \cdot {\bf S}_{j} \rangle
= -\frac{1}{K N_{1}} \frac{1}{\beta}
\frac{\partial \ln Z}{\partial \lambda_{s1}} \nonumber \\
\langle {\bf s}_{1} \cdot {\bf S}_{k} \rangle
&=& \langle {\bf s}_{2} \cdot {\bf S}_{k} \rangle
= -\frac{1}{K^{\prime} N_{3}} \frac{1}{\beta}
\frac{\partial \ln Z}{\partial \lambda_{c1}}
\label{eq:avesS}
\end{eqnarray}
and setting the $\lambda$'s equal to one, this becomes
\begin{equation}
\langle H_{m} \rangle = -\frac{2}{\beta}
\left[ \frac{\partial \ln Z}{\partial \lambda_{s1}}
+ \frac{\partial \ln Z}{\partial \lambda_{c1}} \right]
+ J \langle {\bf s}_{1} \cdot {\bf s}_{2} \rangle .
\label{eq:pairaveHm2}
\end{equation}
Since the true Hamiltonian has the form
\begin{equation}
H = \alpha \sum_{n} {\bf s}_{1} \cdot {\bf S}_{n} | \Psi_{1n} |^{2}
+ \alpha \sum_{n} {\bf s}_{2} \cdot {\bf S}_{n} | \Psi_{2n} |^{2}
+ J {\bf s}_{1} \cdot {\bf s}_{2}
\label{eq:pairH}
\end{equation}
and we know that $\langle {\bf s} \cdot {\bf S}_{n} \rangle$ is
only nonzero for ion spins within the polarons or interstitial region
\begin{eqnarray}
\langle H \rangle
&=& \alpha \langle {\bf s}_{1} \cdot {\bf S}_{i} \rangle \sum_{i} | \Psi_{1i} |^{2}
+ \alpha \langle {\bf s}_{1} \cdot {\bf S}_{k} \rangle \sum_{k} | \Psi_{1k} |^{2}
\nonumber \\
&+& \alpha \langle {\bf s}_{2} \cdot {\bf S}_{j} \rangle \sum_{j} | \Psi_{2j} |^{2}
+ \alpha \langle {\bf s}_{2} \cdot {\bf S}_{k} \rangle \sum_{k} | \Psi_{2k} |^{2}
\nonumber \\
&+& J \langle {\bf s}_{1} \cdot {\bf s}_{2} \rangle .
\label{eq:pairaveH1}
\end{eqnarray}
Again making use of Eq.~(\ref{eq:avesS}), this becomes
\begin{equation}
\langle H \rangle = -\frac{2}{\beta}
\left[ \gamma_{s} \frac{\partial \ln Z}{\partial \lambda_{s1}}
+ \gamma_{c} \frac{\partial \ln Z}{\partial \lambda_{c1}} \right]
+ J \langle {\bf s}_{1} \cdot {\bf s}_{2} \rangle
\label{eq:pairaveH2}
\end{equation}
where $\gamma_{s}$ is defined in Eq.~(\ref{eq:gammasdef}) and we have now
defined
\begin{equation}
\gamma_{c} = \frac{\alpha}{K V_{c}} \int_{C} d^{3}r\, | \Psi_{1} (r) |^2
\label{eq:gammacdef}
\end{equation}
where the integral is over the cylindrically symmetric interstitial region
of volume $V_{c}$.  Subtracting Eq.~(\ref{eq:pairaveHm2}) from
Eq.~(\ref{eq:pairaveH2}) and adding the free energy,
$F_{m}=-\frac{1}{\beta} \ln Z$,
we obtain the polaron-pair ${\cal F}$-function
\begin{equation}
{\cal F} = -\frac{1}{\beta} \left[ \ln Z
+ 2 (\gamma_{s} - 1) \frac{\partial \ln Z}{\partial \lambda_{s1}}
+ 2 (\gamma_{c} - 1) \frac{\partial \ln Z}{\partial \lambda_{c1}} \right]
\label{eq:pairFfunc}
\end{equation}
where $Z$ is the partition function given by Eq.~(\ref{eq:polpair3}).
Since $Z$ is of the form $Z=Z_{1}Z_{2}Z_{3}$, it is clear that ${\cal F}$
is of the form ${\cal F}={\cal F}_{1}+{\cal F}_{2}+{\cal F}_{3}$
and therefore separates into a polaron formation term,
${\cal F}_{1}+{\cal F}_{2}=2{\cal F}_{1}$, which we developed in the
previous section, and a polaron-polaron interaction term
\begin{equation}
{\cal F}_{3} = -\frac{1}{\beta} \left[ \ln Z_{3}
+ 2 (\gamma_{c} - 1) \frac{\partial \ln Z_{3}}{\partial \lambda_{c1}} \right] .
\label{eq:F3def}
\end{equation}
Since these terms share no variational parameters, the polaron-pair model
can be optimized by using ${\cal F}_{1}$ to optimize $K$ and $R$ (as we
did in the previous section) and using ${\cal F}_{3}$ to optimize
$K^{\prime}$ and the parameters describing the geometry of the
interstitial region.

\begin{figure}
\centerline{\psfig{file=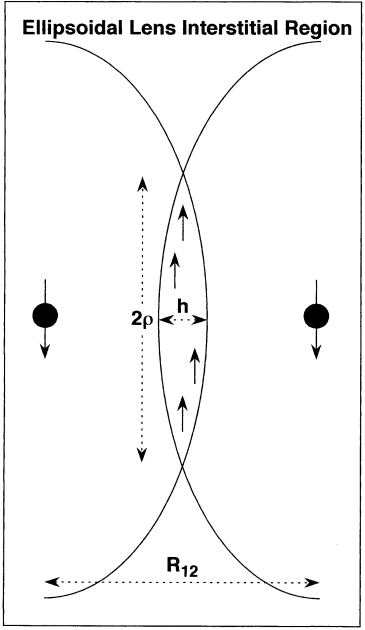}}
\vspace{0.5cm}
\caption{Schematic of ellipsoidal lens shaped interstitial region.}
\label{fig:interstitial}
\end{figure}

Before we can proceed to minimize ${\cal F}_{3}$, we must define a specific
geometry for the interstitial region between the two polarons.  The
bi-spherical geometry of the problem suggests that 
a natural choice would be the spherical lens formed by the intersection of two 
spheres centered on the two polarons.  However, such a shape can be 
completely specified by a single parameter, the lens width $h$.  In order to 
provide an additional degree of freedom within the model, we will use the slightly 
more general ellipsoidal lens formed by the intersection of two ellipsoids centered 
on the polarons.  In this manner, the interstitial region can be specified by two 
parameters, the lens width, $h$, and the lens radius, $\rho$.  The details of
this shape are depicted in Fig.~\ref{fig:interstitial}.

The task of optimizing the model parameters, $K^{\prime}$, $h$, and $\rho$,
is complicated in two ways.  First of all, unlike the single polaron case
where both the partition function, $Z_{1}$, and the geometrical factor,
$\gamma_{s}$, could be obtained analytically, both the interaction
partition function, $Z_{3}$, and the interstitial geometrical factor,
$\gamma_{c}$, must be calculated numerically.  This complicates the
numerics but poses no fundamental problem.

The second complication requires a bit more attention.  Naively
one would expect that by blindly varying these three parameters until ${\cal F}_{3}$
is minimized, the optimal values of these parameters could be obtained.  However, 
upon closer examination, it becomes clear that this is not the case.  In performing 
this optimization of the interstitial region parameters, it is our objective to 
determine the parameters that best match the true interaction between {\it both} 
carriers and the magnetic ions in the interstitial region.  However, minimizing 
the ${\cal F}$-function merely yields the configuration that, overall, is most 
energetically favorable.  As a result, if all three parameters are varied, the 
minimum ${\cal F}_{3}$ will be achieved when the interstitial region
has totally engulfed both of the polarons.
In this configuration, the interstitial region contains ions 
that are very close to the carriers and therefore experience large exchange 
interactions.  However, these ions that interact very strongly with one of the 
polarons barely interact at all with the other.  Thus, the result of an 
unconstrained minimization does not yield an interstitial region with which
{\it both} of the carriers interact strongly.
Therefore, to obtain sensible results, the 
minimization of the interaction ${\cal F}$-function must be constrained.
Two methods of constraining the minimization have been developed: the fixed 
width method and the optimally spherical method.

The interstitial region in which both carrier wave functions are significant 
must be concentrated about the point halfway between the two polarons.  Thus, 
one method of constraining the minimization of the ${\cal F}$-function is to fix 
the lens width, $h$, to a set value.  Using this fixed width method, the only 
parameters that are allowed to vary are the interstitial interaction strength,
$K^{\prime}$, and the lens radius, $\rho$.
For a given temperature, the ${\cal F}$-function will be 
minimized for some optimal values of these two parameters.  Thus, by 
performing a numerical minimization for a range of temperature values, the 
temperature dependence of $K^{\prime}$ and $\rho$ can be determined.

In the fixed width model, the fixed value of $h$ is chosen such that both 
carrier wave functions will be ``significant'' within the interstitial region.  In a 
sense, the choice of a particular $h$ defines the threshold of exchange interaction 
strength with both carriers that is required for an ion to be included in the 
interstitial region.  However, as temperature changes, this threshold should 
change as well since all energies in the system are measured with respect to 
temperature.  Thus, as temperature drops, the threshold should also drop and 
the width of the interstitial region should increase (to include those additional 
ions that now meet the lowered threshold).  This is a feature that is not 
incorporated into the fixed width method, since $h$ remains constant as 
temperature varies.  Unfortunately, this problem cannot be solved by making $h$ 
a variational parameter since, if this is the case, the minimum ${\cal F}$-function 
will only be achieved when the interstitial region engulfs the polarons.  One way 
to incorporate the growth of $h$ with decreasing $T$ into the process would be to 
choose a new fixed $h$ for every $T$ and set the temperature-dependent threshold 
for inclusion in the interstitial region by hand.  However, this approach would 
require a prior knowledge of the temperature dependence of $h$, which we do not 
have.

\begin{figure}
\centerline{\psfig{file=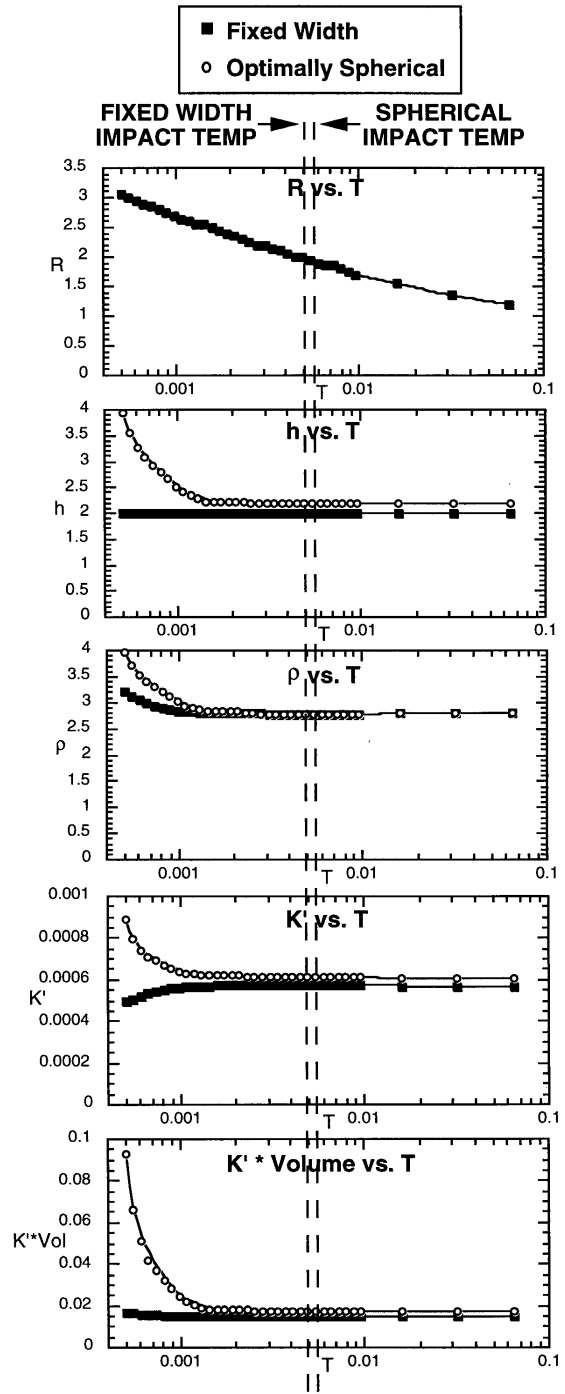}}
\vspace{0.5cm}
\caption{Temperature dependence of polaron-polaron interaction parameters
obtained using both the fixed width and optimally spherical methods of
constraining the minimization.  All distances
are in units of $a_{B}$ and all energies are in units of
$\alpha/a_{B}^{3} \sim 625$~K.}
\label{fig:polpairparams}
\end{figure}

A more natural way to allow the threshold to vary with $T$ is to let the 
threshold be set by the geometry of the system.  The polaron-pair problem 
consists of two sources of radially symmetric wave functions separated by an 
inter-impurity distance.  Thus, the system has an innate bi-spherical geometry.  If 
one were to define a region in space in which the wave functions of both carriers 
would surpass a given threshold, the geometry of the system and the spherically 
symmetric nature of the wave functions would dictate that this region be the 
intersection of two spheres centered at the carrier sites -- a spherical lens.  
Therefore, the natural shape for the interstitial region is a spherical lens.  This 
fact provides the condition for setting the width of the interstitial region in what 
we call the optimally spherical method.  For a given temperature, the 
fixed width, $h$, is set to be that for which the minimization of the
${\cal F}$-function automatically yields a value of the lens radius, $\rho$,
for which the ellipsoidal lens becomes spherical.
Thus, the result is a spherical lens interstitial 
region in which the lens radius is the energetically optimal radius for the given 
width.  As we will show, this technique yields the proper increase in $h$ as 
temperature drops.

Using both the fixed width approach (with $h$ set to $2 a_{B}$)
and the optimally spherical approach,
the polaron-polaron interaction parameters, $K^{\prime}$, $h$, and $\rho$,
were optimized as functions of temperature.  The results of this optimization
for a magnetic ion density of 5 ions per sphere of radius $1a_{B}$ and a
carrier density such that $R_{12}=6a_{B}$ are plotted in
Fig.~\ref{fig:polpairparams}.  Although the parameter values are 
plotted over a wide range of temperatures, they are only meaningful for 
temperatures at which the polaron-pair model is valid.  Clearly, once the 
polarons have increased to a size such that they touch the interstitial region, the 
polaron-pair model is no longer valid.  This impact between the polarons and the 
interstitial region occurs at $T=T_{\mbox{\scriptsize impact}}$ where $2R+h=R_{12}$.
In Fig.~\ref{fig:polpairparams}, the 
temperature of impact has been denoted by a vertical dashed line for both the 
fixed width and the optimally spherical cases.
Note that for either method, $T_{\mbox{\scriptsize impact}}$
is found to be approximately equal to $0.005\, \alpha/a_{B}^{3}$
or around 3~K for typical ion and carrier densities.  Since
$K^{\prime}(T_{\mbox{\scriptsize impact}}) \ll T_{\mbox{\scriptsize impact}}$,
this cutoff temperature lies within the high temperature regime ($T \gg K^{\prime}$)
where the interstitial ion-carrier interactions are insignificant compared to
temperature.  Hence, for temperatures where our
model is well defined, all of the polaron-polaron interaction parameters are 
constant.  For very low temperatures ($T \ll K^{\prime}$), our results are
not quantitative, but since polarons are nearly aligned (either parallel
or antiparallel) by this point, this regime lies beyond the temperature range
of interest.  The use of a constant parameter model to treat the
polaron-polaron interaction in Ref.~\onlinecite{wol96} is therefore justified.

\section{Polaron-Polaron Interaction}
\label{sec:cosine}
\begin{figure}[h]
\centerline{\psfig{file=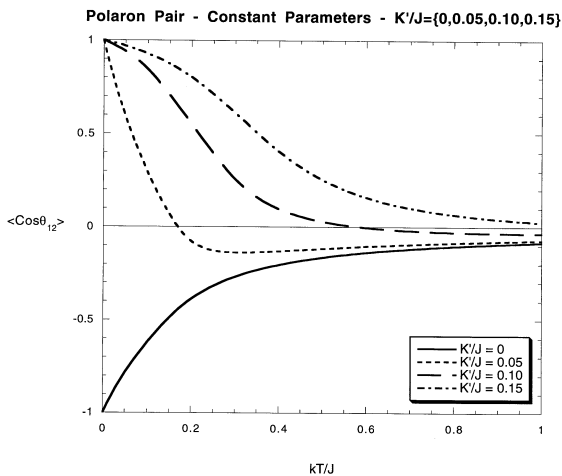}}
\vspace{0.5cm}
\caption{Thermal average of the cosine of the angle between polaron spins,
$\langle\cos\theta_{12}\rangle$, plotted versus temperature for $N_{3}=20$
and several values of the ratio $K^{\prime}/J$.}
\label{fig:interpolangle}
\end{figure}

Given the polaron-polaron interaction partition function, $Z_{3}$,
as expressed in Eq.~(\ref{eq:polpair3}), it follows that
the thermal average of the cosine of the angle between the two polaron spins
can be obtained via
\begin{equation}
\langle \cos \theta_{12} \rangle =
\frac{\int_{0}^{\sqrt{2}} (x^{2} - 1) e^{-\frac{J (x^{2}-1)}{4 k_{B} T}}
F \left( \frac{K^{\prime} x}{\sqrt{2} k_{B} T} \right)^{N_{3}} x\, dx}
{\int_{0}^{\sqrt{2}} e^{-\frac{J (x^{2}-1)}{4 k_{B} T}}
F \left( \frac{K^{\prime} x}{\sqrt{2} k_{B} T} \right)^{N_{3}} x\, dx}
\label{eq:avecosine}
\end{equation}
where $F(z)$ is defined in Eq.~(\ref{eq:Fdef}),
$x^2-1=\cos\theta_{12}$, and $N_{3}=N_{3}(h,\rho)$ is the number of
magnetic ions in the interstitial region.  As motivated above, the
polaron-polaron interaction parameters, $K^{\prime}$ and $N_{3}$,
are approximately temperature-independent in the temperature range
of interest.  Thus, for various constant values of $J$, $K^{\prime}$,
and $N_{3}$, the above expression can be evaluated numerically as a 
function of temperature to reveal the temperature dependence of the
inter-polaron angle.  The results have been plotted in
Fig.~\ref{fig:interpolangle} for $N_{3}=20$ and several values of the ratio
$K^{\prime}/J$.  We see that for reasonable parameter values,
$\langle\cos\theta_{12}\rangle$ is positive and neighboring polarons
tend to align.  For large enough
$K^{\prime}/J$, indirect ferromagnetic carrier-ion-carrier interactions
can dominate the direct antiferromagnetic carrier-carrier interactions
to yield a net ferromagnetic polaron-polaron interaction.

\section{Conclusions}
\label{sec:conclusions}
In this analysis of the magnetic behavior of diluted magnetic 
semiconductors, we have proposed a simplified model to describe both the 
formation of bound magnetic polarons and the interactions between them.
Approximating carrier wave functions via sharp cutoffs that define polarons
and interstitial regions between them, we have obtained a tractable model and
calculated the resulting partition function.
Utilizing a finite temperature variational approach,
the model parameters have been optimized as functions of 
temperature.  At very high temperatures ($T \gg K$), the spins within the system are 
not aligned and the model parameters obtain the constant values for which the 
model wave functions best match the true wave functions.  At lower temperatures
($K^{\prime} \ll T \ll K$), where 
the carrier-ion exchange interaction becomes more significant, both polaron size 
and total intra-polaron exchange energy increase logarithmically with decreasing $T$.
However, throughout this intermediate temperature range, the parameters controlling 
polaron-polaron interactions, $K^{\prime}$, $h$, and $\rho$, remain constant
as $T$ varies.  At very low temperatures ($T \ll K^{\prime}$), even these
interstitial region parameters would become temperature dependent.  However,
for reasonable carrier and ion densities, the polarons grow large enough to 
touch the interstitial region, such that our model ceases to be valid, well
before this regime is realized.  Hence, for temperatures of interest, a model
in which the polaron-polaron interaction parameters are constant is justified
variationally.  Therefore, for reasonable values of $K^{\prime}/J$ we obtain,
as in Ref.~\onlinecite{wol96}, a net ferromagnetic polaron-polaron interaction,
in agreement with the experimental evidence for ferromagnetism in
insulating doped DMS.

The further growth of the polaron pair bubble, for $T \ll K^{\prime}$,
could be captured via a variational scheme employing an appropriately
generalized object.  Starting at the point of polaron overlap, such
an object, involving two or three variational parameters, should evolve
from a ``peanut shape'', with cylindrical symmetry, to a sphere with
the mid-point of the two dopant sites as its center.  This could be
a promising direction for future research.

Since this work was completed, a tendency toward ferromagnetism
has been observed by other groups. \cite{fer00,saw01}
In addition, a separate mechanism for ferromagnetic alignment,
resulting from the strong local exchange fields experienced at the two
impurity sites due to nearby Mn, has been considered by Angelescu and
Bhatt. \cite{ang00}

\acknowledgements
The authors would like to thank NEC Research Institute for allowing
us to reproduce, in Fig.~\ref{fig:doublestep}, the susceptibility data
measured by J. Liu.  This work was supported by NSF DMR-9400362 and 9809483.
ACD acknowledges support from the Princeton Center for Complex Materials
at the Princeton Materials Institute.

\end{document}